\documentclass[slac_one]{revtex4}
\usepackage{graphicx}
\usepackage{fancyhdr}
\begin{document}

\title{Theoretical Overview of Neutrino Properties}

\author{Zhi-zhong Xing}
\affiliation{Institute of High Energy Physics, Chinese Academy of
Sciences, P.O. Box 918, Beijing 100049, China}

\begin{abstract}
I give an overview of some basic properties of massive neutrinos.
The first part of this talk is devoted to three fundamental
questions about three known neutrinos and to their flavor issues
--- the mass spectrum, mixing pattern and CP violation. The second
part of this talk is to highlight a few hot topics at the frontiers
of neutrino physics and neutrino astrophysics, including the
naturalness and testability of TeV seesaw mechanisms at the LHC,
effects of non-standard interactions on neutrino oscillations,
flavor distributions of ultrahigh-energy cosmic neutrinos at
neutrino telescopes, collective flavor oscillations of supernova
neutrinos, flavor effects in thermal leptogenesis, the GSI anomaly
and M$\rm\ddot{o}$ssbauer neutrino oscillations, and so on. I
finally make some concluding remarks for the road ahead.
\end{abstract}

\maketitle

\thispagestyle{fancy}

\section{THE ROAD BEHIND}

The history of neutrino physics can be traced back to the end of
1930, when Pauli wrote that famous letter to ``radioactive ladies
and gentlemen" in T$\rm\ddot{u}$bingen and proposed a desperate
remedy for the energy crisis observed in the beta decay. It is hard
to count how many papers about neutrinos have been published since
then. With the help of the SLAC-SPIRES HEP Database, I have made a
search for papers relevant to neutrinos by inputting ``find title
NEUTRINOS and date XXXX". It turns out that the first neutrino paper
recorded in this HEP archive was the one of Cowan and Reines on the
discovery of $\overline{\nu}^{}_e$ in 1956. Since then, more than
20000 papers on neutrinos have appeared in the literature. Figure 1
shows how the number of papers varies from 1956 to 2007. One can see
some interesting peaks of the curve, which characterize some great
moments in the (incomplete) history of neutrino physics. For
example, the peak around 1968 was triggered by Davis' discovery of
the solar neutrino anomaly; the peak around 1987 was associated with
Koshiba's discovery of the supernova neutrinos; and the peak around
1998 was ascribed to the discovery of atmospheric neutrino anomaly
in the Super-Kamiokande experiment. However, what does the sharp
peak at the beginning of the 1990's mean? I am afraid that this
strange inflation in the number of neutrino papers might be
triggered partly by the 17 keV neutrino episode and partly by the
experimental establishment of $N^{}_\nu = 3$ at the LEP. One may
also identify two golden times in neutrino physics from this
statistical picture: one is the period of 1976 --- 1982 and the
other is from 1998 to the present. I am wondering how long the
second golden time can last.
\begin{figure*}[t]
\centering
\includegraphics[width=135mm]{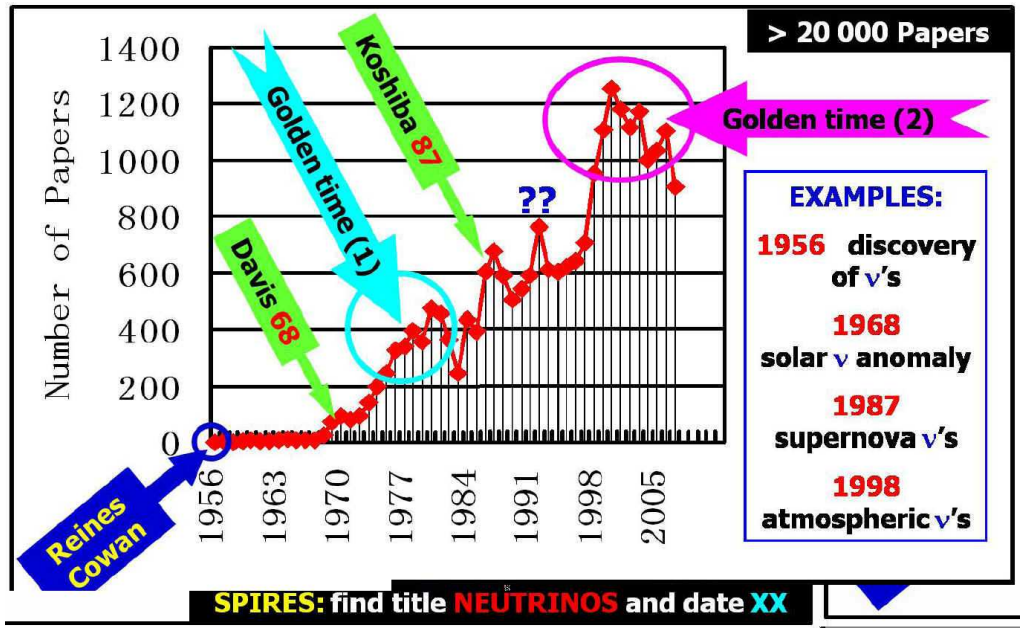}
\caption{A statistical illustration of the history of neutrino
physics and neutrino astrophysics from 1956 to 2007.}
\end{figure*}

The outline of my talk is as follows. In the first part, I am going
to give an overview of some fundamental neutrino properties. They
include the mass puzzle of neutrinos, the nature of massive
neutrinos, the number of neutrino species, and the flavor issues of
neutrino physics (i.e., the mass spectrum, mixing pattern and CP
violation). In the second part of this talk, I shall highlight a few
hot topics at the frontiers of neutrino physics and neutrino
astrophysics, such as the naturalness and testability of TeV seesaw
mechanisms at the LHC, effects of non-standard interactions on
neutrino oscillations, flavor distributions of ultrahigh-energy
cosmic neutrinos at neutrino telescopes, collective flavor
oscillations of supernova neutrinos, flavor effects in thermal
leptogenesis, the GSI anomaly and M$\rm\ddot{o}$ssbauer neutrino
oscillations. Finally, I shall summarize my talk by making some
concluding remarks for the road ahead.

\section{THREE FUNDAMENTAL QUESTIONS}

\subsection{Question 1: Massless or Massive?}

Three known neutrinos ($\nu^{}_e$, $\nu^{}_\mu$, $\nu^{}_\tau$) are
massless in the standard model (SM) as a straightforward consequence
of its simple structure and renormalizability. On the one hand, the
SM does not contain any right-handed neutrinos, and thus there is no
way to write out the Dirac neutrino mass term. On the other hand,
the SM conserves the $SU(2)^{}_L$ gauge symmetry and only contains
the Higgs doublet, and thus the Majorana mass term is forbidden.
Although the SM accidently possesses the $(B-L)$ symmetry and
``naturally" allows neutrinos to be massless, the vanishing of
neutrino masses in the SM is not guaranteed by any {\it fundamental}
symmetry or conservation law. Today we have achieved a lot of robust
evidence for neutrino oscillations from solar, atmospheric, reactor
and accelerator neutrino experiments \cite{Walter}. The phenomenon
of neutrino oscillations implies that at least two neutrinos must be
massive and three neutrino flavors must be mixed. This is the first
convincing evidence for new physics beyond the SM.

\subsection{Question 2: Dirac or Majorana?}

A {\it pure} Dirac mass term added into the SM is theoretically
disfavored, unless the theory is built by introducing extra
dimensions. Such a mass term in the renormalizable models of
electroweak interactions would worsen the problem of large fermion
mass hierarchy; it would violate 't Hooft's naturalness criterion,
as a Majorana mass term of right-handed neutrinos is not forbidden
by the SM gauge symmetry; and it would impose the contrived
assumption of lepton number conservation on the theory. Hence most
theorists believe that massive neutrinos are more likely to be
Majorana particles and their salient feature is lepton number
violation. If massive neutrinos are really Majorana particles, their
masses must have a different origin in comparison with the masses of
charged leptons and quarks.

The only experimentally feasible way to verify the Majorana nature
of massive neutrinos is to observe the neutrinoless doube-beta
($0\nu \beta\beta$) decay. So far we have not obtained very
convincing evidence for this lepton-number-violating process. Note
that an uncontrovertible observation of the $0\nu \beta\beta$ decay
will definitely imply that massive neutrinos are Majorana particles
\cite{Valle}, but it may not uniquely point to neutrino masses and
flavor mixing quantities.

It is worth emphasizing that a massive Dirac neutrino, given the
SM interactions, can have a tiny (one-loop) magnetic dipole moment
$\mu^{}_\nu \sim 3\times 10^{-20} \mu^{}_{\rm B} (m^{}_\nu/0.1 ~
{\rm eV})$, where $\mu^{}_{\rm B}$ is the Bohr magneton
\cite{Shrock}. In contrast, a massive Majorana neutrino cannot
have magnetic and electric dipole moments, because its
antiparticle is just itself. Both Dirac and Majorana neutrinos can
have {\it transition} dipole moments (of a size comparable with
$\mu^{}_\nu$) \cite{Vissani}, which may give rise to neutrino
decays; scattering effects with electrons; interactions with
external magnetic fields (red-giant stars, the sun, supernovae,
and so on); and contributions to neutrino masses. Current
experimental bounds on neutrino dipole moments are at the level of
$\mu^{}_\nu < {\rm a ~ few} \times 10^{-11} \mu^{}_{\rm B}$.

\subsection{Question 3: Three Species or More?}

It is well known that ``three" is a mystically popular number in
particle physics: three $Q=+2/3$ quarks; three $Q=-1/3$ quarks;
three $Q=-1$ leptons; three $Q=0$ neutrinos; three colors; and three
forces in the SM. In this case, why do not we just consider three
species of neutrinos and why do we consider to go beyond $N^{}_\nu =
3$?

In the past one and a half decades, the main motivation for some
theorists to speculate the existence of light {\it sterile}
neutrinos was to account for the LSND anomaly together with solar
and atmospheric neutrino oscillations. A global analysis of current
neutrino oscillation data disfavors plain (3+1), (3+2) and even
(3+3) scenarios of active-sterile neutrino mixing \cite{Maltoni}.
The recent MiniBOONE experiment does not support the LSND result
either. Conservatively speaking, it would be a big surprise if such
exotic particles were really staying with us and in the universe.

A very large number of theorists are motivated by the elegant
(type-I) seesaw mechanism to consider the existence of heavy
Majorana neutrinos, because it is currently the most natural way to
understand the origin of neutrino masses and why they are so tiny.
If three known neutrinos really have three unknown partners whose
masses are above or far above the Fermi scale, an exciting window
will be open to new physics at high energy scales. In this case,
however, the mixing between light and heavy neutrinos violates the
unitarity of the $3\times 3$ light neutrino mixing matrix and might
result in some observable effects in the future precision neutrino
oscillation experiments.

\section{ISSUES OF NEUTRINO FLAVORS}

There are three central concepts in flavor physics: mass, flavor
mixing and CP violation \cite{Xing04}. Fogli {\it et al}
\cite{Fogli} have recently done a global analysis of current
neutrino oscillation data and obtained the ranges of two neutrino
mass-squared differences ($\delta m^2 \equiv m^2_2 - m^2_1$ and
$\Delta m^2 \equiv |m^2_3 - (m^2_1 + m^2_2)/2|$) and three neutrino
mixing angles ($\theta^{}_{12}$, $\theta^{}_{13}$ and
$\theta^{}_{23}$ in the standard parametrization of the $3\times 3$
unitary neutrino mixing matrix $V$ \cite{PDG08}), as listed in Table
I.
\begin{table}[ph]
\begin{center}
\caption{The latest global 3$\nu$-oscillation analysis
\cite{Fogli}.}
\begin{tabular}{c|c|c|c|c|c}
\toprule Parameter & $\delta m^2/10^{-5} ~ {\rm eV}^2$ &
$\sin^2\theta^{}_{12}$ & $\sin^2\theta^{}_{13}$ &
$\sin^2\theta^{}_{23}$ & $\Delta m^2/10^{-3} ~ {\rm eV}^2$ \\
\colrule
Best fit & 7.67 & 0.312 & 0.016 & 0.466 & 2.39 \\
1$\sigma$ range & ~ 7.48$\cdots$7.83 ~ & ~ 0.294$\cdots$0.331 ~ &
~ 0.006$\cdots$0.026 ~ & ~ 0.408$\cdots$0.539 ~ & ~ 2.31$\cdots$2.50 ~ \\
2$\sigma$ range & 7.31$\cdots$8.01 & 0.278$\cdots$0.352 & $<$ 0.036
& 0.366$\cdots$0.602 & 2.19$\cdots$2.66 \\
3$\sigma$ range & 7.14$\cdots$8.19 & 0.263$\cdots$0.375 & $<$ 0.046
& 0.331$\cdots$0.644 & 2.06$\cdots$2.81 \\ \botrule
\end{tabular}
\end{center}
\end{table}
\vspace{-0.5cm}

\subsection{Neutrino Mass Spectrum}

Two mass-squared differences of three known neutrinos have been
determined, to a good degree of accuracy, from current
experimental data: $\Delta m^2_{21} \approx 7.7 \times 10^{-5} ~
{\rm eV}^2$ and $\Delta m^2_{32} \approx \pm 2.4 \times 10^{-3} ~
{\rm eV}^2$. The absolute neutrino mass scale remains unknown and
may hopefully be determined in three experimental or observational
ways: the single beta decay; the $0\nu \beta\beta$ decay; and the
cosmological constraints. Considering the tight and loose
constraints from cosmology together with the present neutrino
oscillation results, Fogli {\it et al} \cite{Fogli} have analyzed
the parameter space of $\sum m^{}_i$ and $m^{}_{\beta\beta}$ as
shown in Figure 2, where $m^{}_{\beta\beta} = \sum \left( m^{}_i
V^2_{ei} \right)$ is the effective mass of the $0\nu \beta\beta$
decay. One can see that the Heidelberg-Moscow claim for the $0\nu
\beta\beta$ evidence \cite{HM} is compatible with the loose CMB
data and, if finally confirmed by other experiments, would imply a
near degeneracy of three neutrino masses.
\begin{figure*}[t]
\centering
\includegraphics[width=135mm]{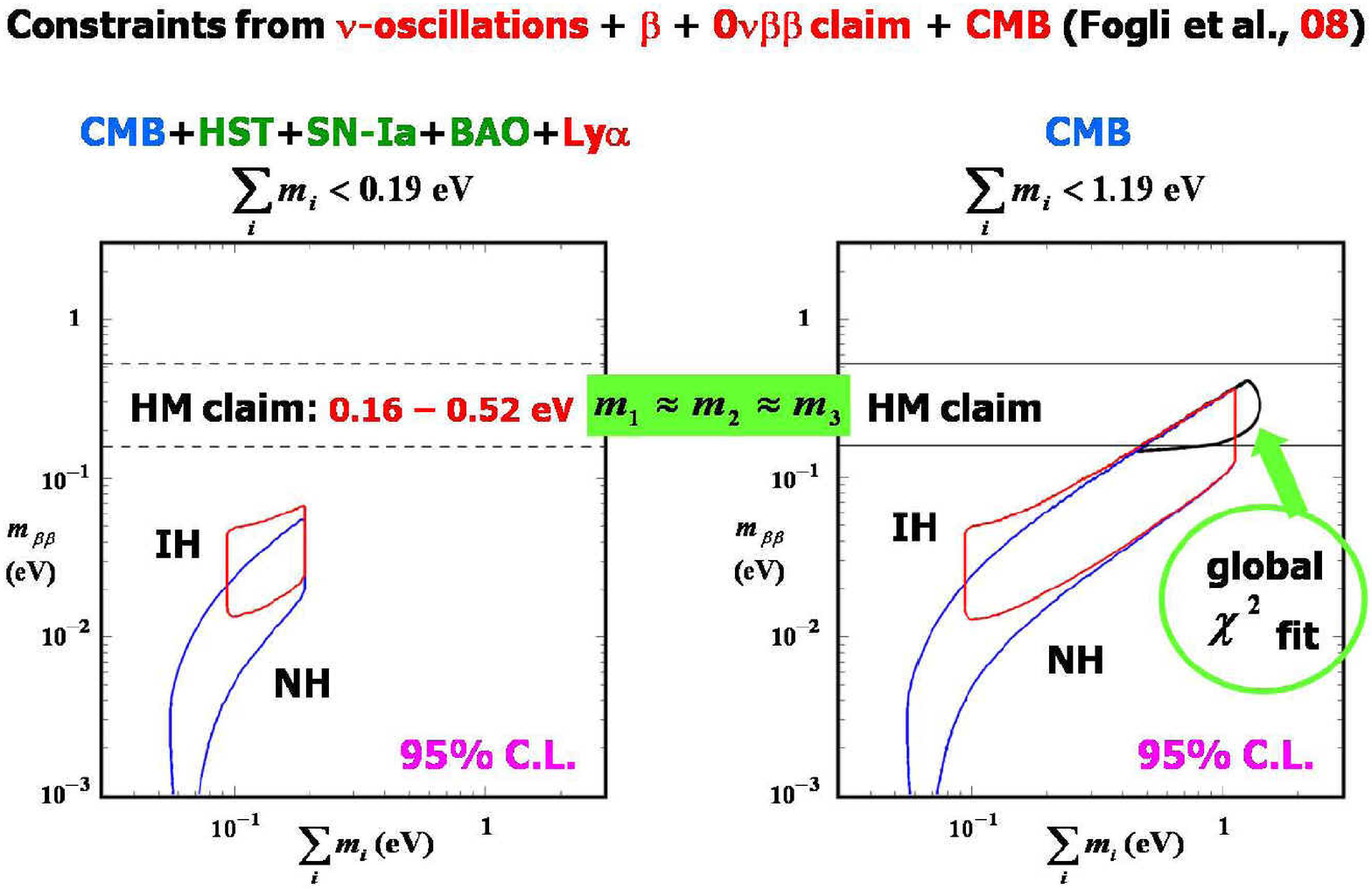}
\caption{Bounds on $\sum m^{}_i$ and $m^{}_{\beta\beta}$ from
cosmological and experimental neutrino data.}
\end{figure*}

Before the absolute neutrino mass scale is fixed, there remain two
open questions: (1) is $m^{}_3$ bigger or smaller than $m^{}_1$
(i.e., normal or inverted hierarchy)? (2) can one neutrino mass
($m^{}_1$ or $m^{}_3$) be vanishing or vanishingly small? Question
(1) requires an experimental answer in the near future, such as the
long-baseline neutrino oscillation experiments with appreciable
terrestrial matter effects \cite{GG}; and question (2) depends on
the special structure or symmetry of a realistic neutrino mass
model, such as the minimal type-I seesaw mechanism with two heavy
Majorana neutrinos \cite{MSM} or the Friedberg-Lee symmetry of an
effective Dirac or Majorana neutrino mass operator \cite{FL}.

\subsection{Flavor Mixing Pattern}

In the standard neutrino phenomenology, the flavor mixing matrix of
three neutrinos $V$ is assumed to be {\it unitary} and can be
parametrized in terms of three rotation angles ($\theta^{}_{12}$,
$\theta^{}_{13}$, $\theta^{}_{23}$) and three CP-violating phases
($\delta$, $\rho$, $\sigma$). Current experimental constraints on
$\theta^{}_{12}$, $\theta^{}_{13}$ and $\theta^{}_{23}$ are given in
Table I, but $\delta$, $\rho$ and $\sigma$ are entirely
unrestricted. The upper bound on $\theta^{}_{13}$ is $\theta^{}_{13}
< \theta^{}_{\rm C} \approx 13^\circ$ at the $3\sigma$ level, and a
global analysis of all neutrino oscillation data yields
$\sin^2\theta^{}_{13} = 0.016 \pm 0.010$ at the $1\sigma$ level
\cite{Fogli}. How small $\theta^{}_{13}$ is remains an open
question. Reactor and accelerator neutrino oscillation experiments
will hopefully answer this question in a direct way, but can they
answer it before the global fit indirectly ``predicts" the value of
$\theta^{}_{13}$?

We see that $\theta^{}_{12}$ and $\theta^{}_{23}$ are both large and
close to two special numbers: $\theta^{}_{12} \sim
\arctan(1/\sqrt{2}) \approx 35.3^\circ$ and $\theta^{}_{23} \sim
45^\circ$. This naive observation implies that the realistic
neutrino mixing pattern might result from a certain underlying
flavor symmetry (e.g., $S^{}_3$, $S^{}_4$, $A^{}_4$, $Z^{}_2$,
$U(1)^{}_{\rm F}$, $\cdots$ \cite{Feruglio}) and its spontaneous or
explicit breaking. One may play games with a few small integers and
their square roots to reconstruct the form of $V$ in an economical
group language. A typical example is the so-called tri-bimaximal
neutrino mixing pattern \cite{TB}, whose three angles happen to be
$\theta^{}_{12} = \arctan(1/\sqrt{2}) \approx 35.3^\circ$,
$\theta^{}_{13} = 0^\circ$ and $\theta^{}_{23} = 45^\circ$. So far
many interesting possibilities of model building have been tried
with the help of many flavor symmetries \cite{Review}, and this
situation seems to be promising on the one hand and giving rise to
the uniqueness problem on the other hand.

\subsection{CP and T Violation}

If neutrinos are Majorana particles, the $3\times 3$ unitary
neutrino mixing matrix $V$ contains three CP-violating phases
$\delta$, $\rho$ and $\sigma$. Among them, $\delta$ determines the
strength of CP and T violation in neutrino oscillations, because
both $P(\nu^{}_\alpha \to \nu^{}_\beta) - P(\overline{\nu}^{}_\alpha
\to \overline{\nu}^{}_\beta)$ and $P(\nu^{}_\alpha \to \nu^{}_\beta)
- P(\nu^{}_\beta \to \nu^{}_\alpha)$ are proportional to the
Jarlskog invariant ${\cal J} = \sin\theta^{}_{12} \cos\theta^{}_{12}
\sin\theta^{}_{23} \cos\theta^{}_{23} \sin\theta^{}_{13}
\cos^2\theta^{}_{13} \sin\delta$ in vacuum. The Majorana phases
$\rho$ and $\delta$, which have nothing to do with neutrino
oscillations, are associated with the $0\nu \beta\beta$ decay. Note
that $\delta$ itself is also of the Majorana nature, although it is
usually referred to as the Dirac phase: one reason is that $\delta$
may appear in other lepton-number-violating processes, even if it
can always be arranged {\it not} to appear in the $0\nu \beta\beta$
decay; and the other reason is that $\delta$, $\rho$ and $\sigma$
are actually entangled with one another in the renormalization-group
running from one energy scale to another \cite{RGE}.

It is worth pointing out that ${\cal J}$ takes its maximal value
${\cal J}^{}_{\rm max} = 1/(6\sqrt{3}) \approx 9.6\%$ when
$\theta^{}_{12} = \theta^{}_{23} = 45^\circ$, $\theta^{}_{13} =
\arctan(1/\sqrt{2}) \approx 35.3^\circ$ and $\delta = 90^\circ$
hold. Here again the mysterious value $35.3^\circ$ shows up. Note
that such a special angle has a simple geometric explanation
\cite{Lee}: it corresponds to the angle formed by two unequal
diagonals from the same vertex of a cube, and thus it might imply a
certain flavor symmetry described by a simple group language. If
$\delta$ is large and $\theta^{}_{13}$ is not too small, it will be
hopeful to observe leptonic CP and T violation in the future
long-baseline neutrino oscillation experiments. Of course,
terrestrial matter effects, which might fake the genuine signals of
CP or T violation, have to be taken into account in a realistic
long-baseline experiment.

Taking account of the lesson that we have learnt from the
phenomenology of quark flavor mixing and CP violation, I stress that
a known value of the smallest neutrino mixing angle $\theta^{}_{13}$
will be an important turning point to the era of precision
measurements in experimental neutrino physics, towards the
observation of CP violation, the test of unitarity of $V$, and the
search for possible new (expected or unexpected) physics in the long
run.

\section{SELECTED HOT TOPICS}

\subsection{TeV Seesaws and Collider Signatures}

The origin of finite but tiny neutrino masses is a big puzzle in
particle physics. Among many theoretical and phenomenological ideas
to solve this problem, the seesaw picture seems to be most natural
and elegant at present. Three typical seesaw mechanisms are
illustrated in Figure 3, and some other variations or combinations
are possible.
\begin{figure*}[t]
\centering
\includegraphics[width=135mm]{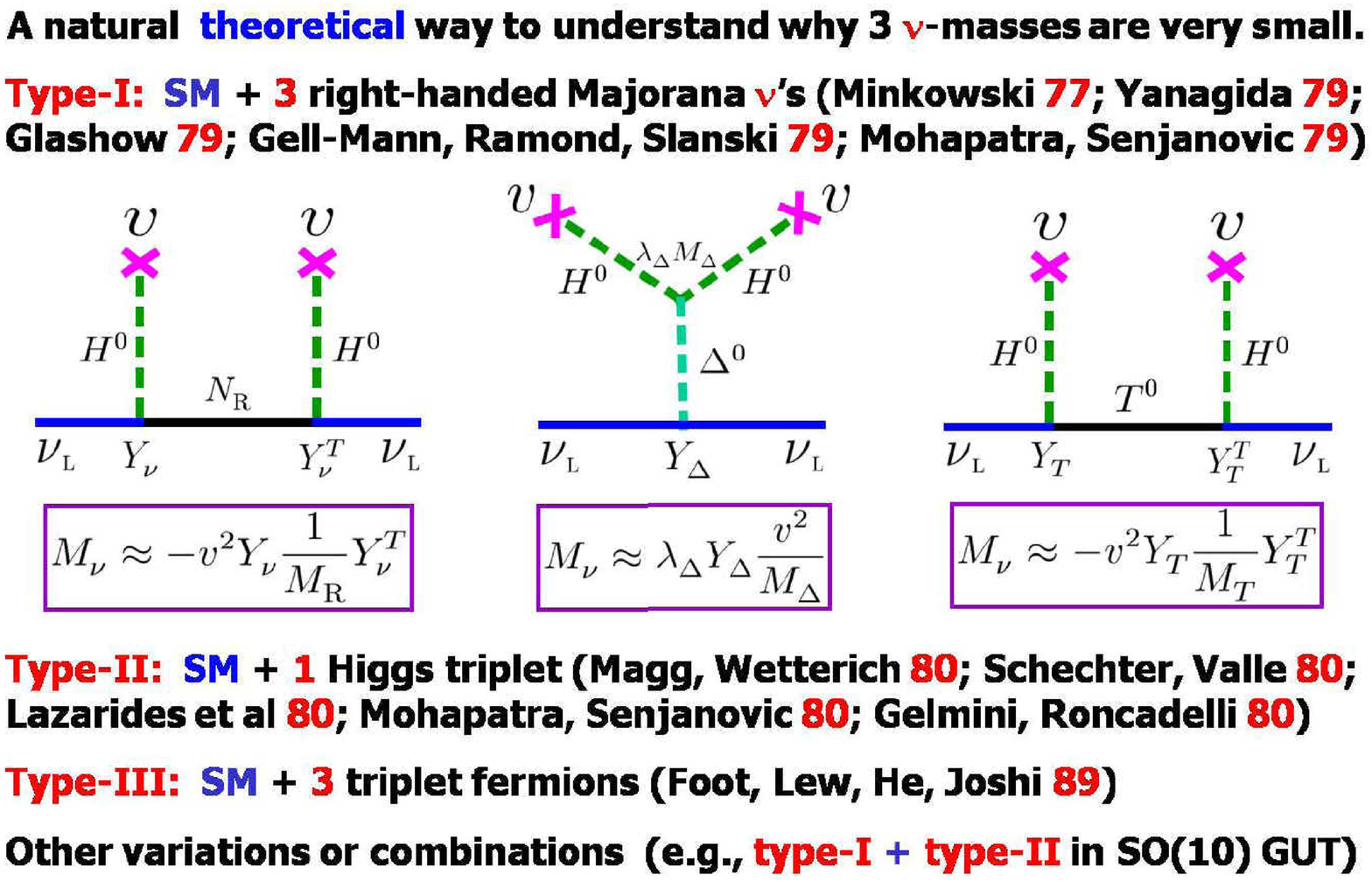}
\caption{Three types of seesaw mechanisms to understand finite but
tiny neutrino masses.}
\end{figure*}

The scale where a seesaw mechanism works is crucial, because it is
relevant to whether this mechanism is theoretically natural and
experimentally testable. Between Fermi and Planck scales, there
might exist two other fundamental scales: one is the GUT scale at
which strong, weak and electromagnetic forces can be unified, and
the other is the TeV scale at which the unnatural gauge hierarchy
problem of the SM can be solved or softened by new physics. Many
theorists argue that the conventional seesaw pictures are natural
because their scales (i.e., the masses of heavy degrees of freedom)
are close to the GUT scale. If the TeV scale is also a fundamental
scale, can we argue that the TeV seesaws are also natural? In other
words, we are reasonably motivated to speculate that possible new
physics existing at the TeV scale and responsible for the
electroweak symmetry breaking might also be responsible for the
origin of neutrino masses. It is therefore interesting and
meaningful to balance the ``naturalness" and ``testability" of TeV
seesaws at the energy frontier set by the LHC.

Among many works done in the past two years about TeV seesaws (see
Refs. \cite{S1,S2,S3} for an incomplete list), at least the
following lessons can be learnt. First, lepton-number-violating
($\Delta L =2$) like-sign dilepton events are clean collider
signatures of heavy seesaw particles in most cases \cite{KS}.
Typical signatures at the LHC include $pp \to W^{*\pm} \to
l^\pm_\alpha N \to l^\pm_\alpha l^\pm_\beta jj$ (type-I seesaw);
$pp \to \gamma^*, Z^* \to H^{++}H^{--}$ and $pp \to W^{*\pm} \to
H^{\pm\pm} H^{\mp}$ with $H^{\pm\pm} \to l^\pm_\alpha l^\pm_\beta$
(type-II seesaw); and $pp \to W^{*\pm} \to T^\pm T^0 \to
l^\pm_\alpha l^\pm_\beta + ZW^\mp (\to 4 j)$ (type-III seesaw).
Second, the LHC signatures of heavy Majorana neutrinos and triplet
fermions are likely to be decoupled from the parameters of light
Majorana neutrino masses. Third, the level of fine-tuning in TeV
seesaws to get a mass scale of $0.1$ eV for light Majorana
neutrinos could even be $10^{-10}$
--- very unnatural? In addition, the mixing between light
neutrinos and heavy particles in type-I and type-III seesaws may
lead to observable unitarity violation of the $3\times 3$ light
neutrino mixing matrix \cite{Xing08}.

\subsection{Non-standard Interactions and Non-unitary Neutrino Oscillations}

Non-standard interactions (NSIs) of neutrinos can be described by
the effective four-fermion operators, ${\cal L}^{}_{\rm NSI} \approx
2\sqrt{2} ~G^{}_{\rm F} (\overline{\nu^{}_{\beta \rm L}} ~\gamma^\mu
\nu^{}_{\alpha \rm L}) (\overline{f^{}_{\rm L}} ~\gamma^{}_\mu
f^{}_{\rm L}) \epsilon^{}_{\alpha \beta}$ at low energy scales,
after heavy degrees of freedom are integrated out. The magnitude of
$\epsilon^{}_{\alpha\beta}$ is expected to be
$|\epsilon^{}_{\alpha\beta}| \sim M^2_W/M^2_{\rm NSI} \lesssim {\cal
O}(0.1)$, as constrained by current experimental data. The effects
of NSIs can show up not only in the $0 \nu \beta\beta$ decay and
lepton-flavor-violating rare processes but also in neutrino
oscillations (at the source, at the detector and in propagation)
\cite{Lindner0}. Future precision neutrino oscillation experiments
might be able to probe such sub-leading new physics effects beyond
the SM \cite{NSI}, e.g., through the measurements of interference
terms in neutrino oscillations and new CP-violating phenomena.

The {\it minimal unitarity violation} in the neutrino sector can be
regarded as a special NSI of neutrinos at low energies. In this
scheme, only three light neutrino species are considered, and the
sources of non-unitarity are allowed only in those terms of the SM
Lagrangian which involve neutrinos \cite{Antusch}. TeV seesaws are
therefore a good framework which can naturally accommodate
non-unitary mixing of three light neutrinos. A careful analysis of
current experimental data shows that the unitarity of the $3\times
3$ light neutrino mixing matrix $V$ is good at the percent level,
implying that possible non-unitary effects should at most be at the
percent level. A salient feature of non-unitary neutrino
oscillations is the ``zero-distance" (near detector) effect:
$P(\nu^{}_\alpha \to \nu^{}_\beta)|^{}_{L=0} =
|(VV^\dagger)^{}_{\alpha\beta}|^2
/[(VV^\dagger)^{}_{\alpha\alpha}(VV^\dagger)^{}_{\beta\beta}] \neq
0$ for $\alpha \neq \beta$; i.e., a flavor transition can take place
even at the source before neutrino oscillations really develop.
Another interesting consequence of non-unitary neutrino mixing is
the new CP-violating effect in short- or medium-baseline $\nu^{}_\mu
\to \nu^{}_\tau$ and $\overline{\nu}^{}_\mu \to
\overline{\nu}^{}_\tau$ oscillations \cite{Yasuda}. We find
$P(\overline{\nu}^{}_\mu \to \overline{\nu}^{}_\tau) - P(\nu^{}_\mu
\to \nu^{}_\tau) \approx 2 \sin\theta^{}_{24} \sin\theta^{}_{34}
\sin (\delta^{}_{24} - \delta^{}_{34}) \sin (0.5 \Delta m^2_{32}
L/E)$ in vacuum, where $\theta^{}_{ij}$ and $\delta^{}_{ij}$
describe the new rotation angles and CP-violating phases due to the
mixing between three light Majorana neutrinos and one heavy Majorana
neutrino in a simplified seesaw scenario \cite{Xing08}. This
non-trivial signature of CP violation can maximally be at the
percent level, and it might be contaminated by terrestrial matter
effects if the baseline length is sufficiently long \cite{Luo}.

\subsection{Flavor Effects in Thermal Leptogenesis}

Fukugita and Yanagida's canonical idea of baryogenesis via
leptogeneis \cite{FY} works in the type-I seesaw mechanism with two
or more heavy Majorana neutrinos, where lepton number is violated at
the tree level and direct CP violation occurs at the one-loop level
of heavy Majorana neutrino decays. Two key points of thermal
leptogenesis: (1) the CP-violating asymmetry between the
out-of-equilibrium decay of the lightest heavy Majorana neutrino and
its CP-conjugate process is partly converted into a net lepton
number asymmetry; (2) the latter is finally converted into a net
baryon number asymmetry via $(B-L)$-conserving but $(B+L)$-violating
sphaleron processes, and thus the observed baryon number asymmetry
of the Universe (i.e., $\eta^{}_{\rm B} \equiv n^{}_{\rm
B}/n^{}_\gamma \approx 6.1 \times 10^{-10}$ \cite{PDG08}) can
naturally be interpreted.

In their latest review paper \cite{Nir1}, Davidson {\it et al} have
given a comprehensive summary of recent developments in thermal
leptogenesis, including the studies of spectator processes, finite
temperature effects, flavor effects, $N^{}_2$ leptongenesis and
resonant (TeV) leptogenesis. Some variations of the conventional
ideas have also been discussed in the literature \cite{Nir2}: soft
(SUSY) leptogenesis, type-II leptogenesis, type-III leptogenesis,
Dirac leptogenesis, electromagnetic leptogenesis, and so on. Here
let me briefly comment on the flavor effects in thermal
leptogenesis.
\begin{figure*}[t]
\centering
\includegraphics[width=135mm]{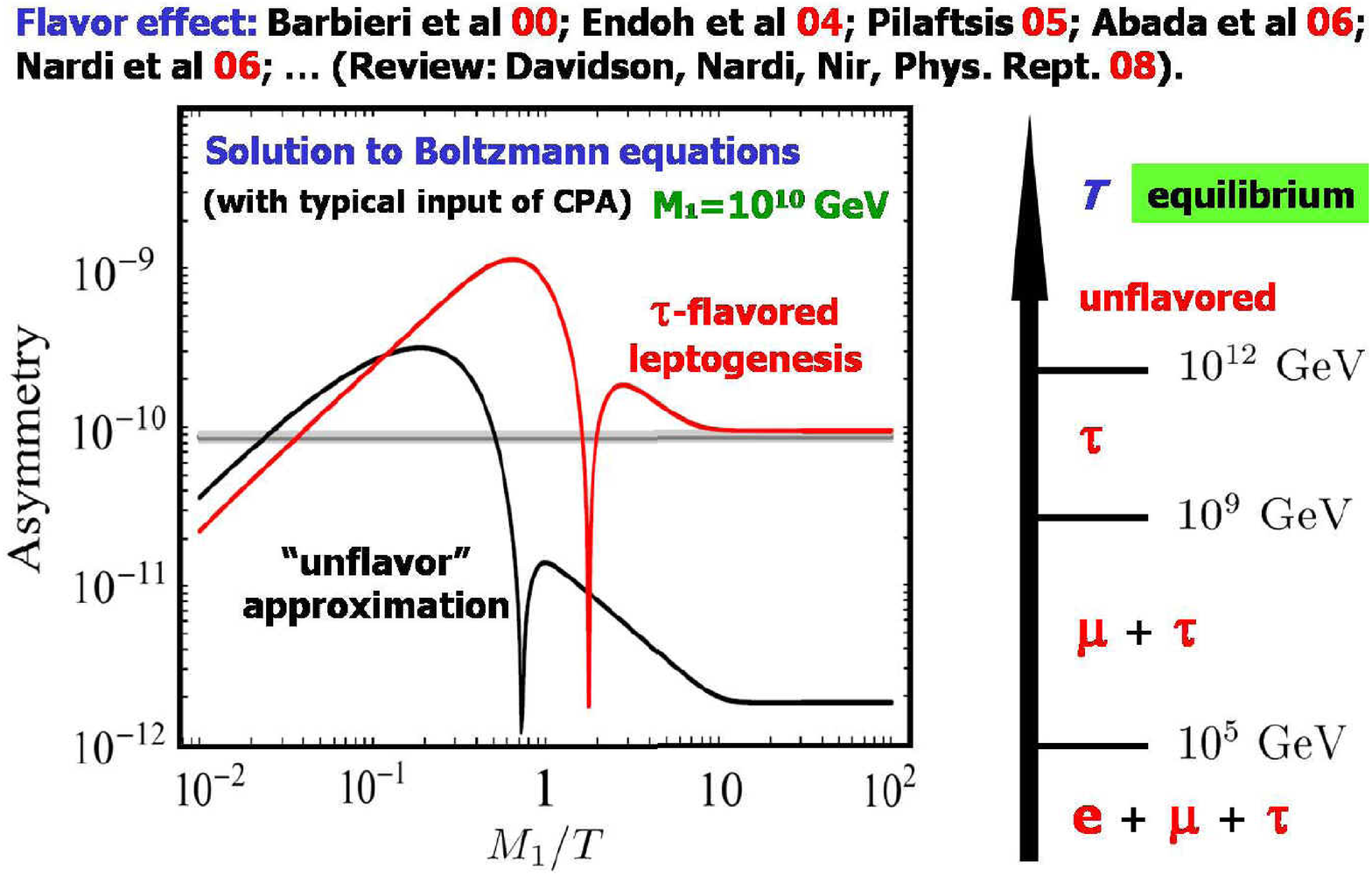}
\caption{A simple illustration of possibly important flavor effects
in thermal leptogenesis.}
\end{figure*}

Figure 4 illustrates why flavor effects might be important in
leptogenesis. One can see that four temperature intervals are of
interest in solving the Boltzmann equations which describe the
evolution of the baryon number asymmetry \cite{Barbieri}: (1) above
$T \sim 10^{12}$ GeV, the leptonic Yukawa interactions are not in
equilibrium --- ``unflavored" leptogenesis; (2) between $T \sim
10^9$ GeV and $T \sim 10^{12}$ GeV, the $\tau$-lepton Yukawa
interactions are in equilibrium --- $\tau$-flavored leptogenesis;
(3) between $T \sim 10^5$ GeV and $T \sim 10^9$ GeV, both $\mu$- and
$\tau$-lepton Yukawa interactions are in equilibrium --- $(\mu +
\tau)$-flavored leptogenesis; and (4) below $T \sim 10^5$ GeV, all
the leptonic Yukawa interactions are in equilibrium ---
fully-flavored leptogenesis. A lot of interesting works on flavor
effects in thermal leptogenesis have been done since 2004
\cite{Flavor}. The most striking consequence of such effects might
be that it is possible to establish a direct relationship between
the cosmological baryon number asymmetry and the CP violation at low
energies via flavored leptogenesis in a class of seesaw models.

\subsection{Flavor Distributions of UHE Cosmic Neutrinos}

Now that neutrinos can oscillate from one flavor to another, it will
be extremely interesting to detect the oscillatory phenomena of
ultrahigh-energy (UHE) cosmic neutrinos produced from distant
astrophysical sources. IceCube \cite{IceCube}, a ${\rm km}^3$-volume
under-ice neutrino telescope, is now under construction at the South
Pole and aims to observe the UHE neutrino oscillations. Together
with the under-water neutrino telescopes in the Mediterranean Sea
\cite{Sea}, IceCube has the potential to shed light on the
acceleration mechanism of UHE cosmic rays and to probe the intrinsic
properties of cosmic neutrinos. An immediate consequence of neutrino
oscillations is that the flavor composition of cosmic neutrinos to
be observed at the telescopes must be different from that at the
sources \cite{Pakvasa}. By measuring the cosmic neutrino flavor
distribution, one can either constrain the standard neutrino mixing
parameters ($\theta^{}_{12}$, $\theta^{}_{13}$, $\theta^{}_{23}$ and
$\delta$) or probe possible new physics beyond the SM (e.g., CPT
violation, quantum decoherence, unitarity violation, neutrino
decays, and active-sterile neutrino oscillations). A lot of
attention has been paid to these possibilities
\cite{Serpico,Xing06,NT,Decay}.

Let me make two brief remarks \cite{Xing06}. (1) For the
conventional sources of UHE cosmic neutrinos (produced from the
decays of charged pions arising from UHE $pp$ and (or) $p\gamma$
collisions) with $\phi^{}_e : \phi^{}_\mu : \phi^{}_\tau = 1 : 2 :
0$, we can get the democratic flavor distribution $\phi^{\rm T}_e :
\phi^{\rm T }_\mu : \phi^{\rm T}_\tau = 1 : 1 : 1$ at a terrestrial
neutrino telescope under the condition $|V^{}_{\mu i}| = |V^{}_{\tau
i}|$ (for $i=1, 2, 3$), which is equivalent to either
$\theta^{}_{13} =0$ and $\theta^{}_{23} = \pi/4$ (CP-conserving
case) or $\delta = \pm \pi/2$ and $\theta^{}_{23} = \pi/4$
(CP-violating case). (2) If there exist some contaminations to the
conventional sources of UHE cosmic neutrinos or if the astrophysical
sources are quite different from the conventional ones, one may
adopt a generic parametrization of neutrino flavors at the sources:
$\phi^{}_e : \phi^{}_\mu : \phi^{}_\tau = \sin^2\xi \cos^2\zeta :
\cos^2\xi \cos^2\zeta : \sin^2\zeta$. Both $\xi$ and $\zeta$ can in
principle be measured at neutrino telescopes, although this is
extremely difficult (if not impossible) in practice.

\subsection{Collective Neutrino Flavor Transitions in Supernovae}

Neutrinos streaming off a collapsed supernova core are so dense near
the neutrino sphere that their nonlinear self-interactions are very
significant \cite{Fuller} and can give rise to {\it collective}
flavor transitions \cite{Collective1} in which neutrinos (or
antineutrinos) of different energies almost have the same behaviors.
The practical importance of such effects has recently been
recognized and has triggered some intensive studies
\cite{Collective2}.

The condition for neutrino-neutrino scattering effects to be
relevant can simply be explained in a two-flavor neutrino
oscillation scenario, where $\omega = \Delta m^2/(2E)$ is the vacuum
oscillation frequency, $\lambda = \sqrt{2} G^{}_{\rm F} n^{}_e$
describes the ordinary MSW matter effects, and $\mu = \sqrt{2}
G^{}_{\rm F} (n^{}_\nu + n^{}_{\overline{\nu}})$ denotes the
self-interactions of neutrinos and antineutrinos. When $\omega
\lesssim \lambda$, the MSW matter effects are important; and when
$\omega \lesssim \mu$, neutrino-neutrino scattering effects are
important. Note that it is a misconception that neutrino-neutrino
scattering effects would be negligible even if $\mu \ll \lambda$
holds.

Some interesting properties of collective supernova neutrino
oscillations have been discussed in the literature
\cite{Collective2}. {\it Synchronized oscillations} take place
when the self-interactions of neutrinos and antineutrinos ``glue"
the neutrino flavor polarization vectors together such that they
evolve in the same way. {\it Bipolar oscillations} occur in a
neutrino gas with equal densities of neutrinos and antineutrinos
(e.g., $\nu^{}_e$ and $\overline{\nu}^{}_e$). For the inverted
neutrino mass hierarchy with very small $\theta^{}_{13}$, the
ensemble will undergo oscillations of the type $\nu^{}_e
\overline{\nu}^{}_e \to \nu^{}_\mu \overline{\nu}^{}_\mu \to
\nu^{}_e \overline{\nu}^{}_e \to \cdots$, approximately with the
``bipolar frequency" $\kappa = \sqrt{2\omega \mu} \gg \lambda$.
For the normal neutrino mass hierarchy, the ensemble will perform
small-amplitude harmonic oscillations with the frequency $\kappa$,
and thus nothing happens macroscopically. It has also been found
that the collective effects can lead to a {\it spectral split} or
{\it stepwise swap} of neutrino flavors, where a critical energy
splits the transformed spectrum sharply into parts of almost pure
but different flavors \cite{Raffelt}.

\subsection{The GSI Anomaly and M$\bf\ddot{\bf o}$ssbauer Neutrino Oscillations}

The orbital electron capture decays of hydrogen-like $^{140}{\rm
Pr}$ and $^{142}{\rm Pm}$ ions, similar to the $p + e^- \to n +
\nu^{}_e$ process, have recently been measured at GSI Darmstadt by
using the time-resolved Schottky mass spectrometry \cite{GSI}. The
experimental result, contrary to the expected {\it pure}
exponential behavior, can be described by an exponential fit plus
a superimposed oscillation at the $3.5\sigma$ level. Several
authors have argued that this anomaly might be attributed to
neutrino mixing in the final state \cite{Lipkin}, but their
arguments seem to be unconvincing. As carefully analyzed by some
other authors \cite{Giunti}, the GSI anomaly cannot originate from
neutrino mixing. If it finally survives, it might be associated
with the properties of $^{140}{\rm Pr}$ and $^{142}{\rm Pm}$ ions
themselves.

Some special interest has recently been paid to the possibility of
measuring $\theta^{}_{13}$, the smallest neutrino mixing angle, by
doing a M$\rm\ddot{o}$ssbauer neutrino oscillation experiment
\cite{Raghavan}. The basic idea of such an experiment is rather
simple, as illustrated in Figure 5. If $^3{\rm H}$ and $^3{\rm
He}$ are both embedded into a solid-state lattice (e.g., metal
crystals), the recoilless emission and resonant absorption of
$\overline{\nu}^{}_e$ neutrinos will in principle be possible
\cite{Visscher}. We refer to the nearly monochromatic
$\overline{\nu}^{}_e$ beam with energy $E=18.6$ keV in these
reactions as the M$\rm\ddot{o}$ssbauer neutrinos, because the
mechanism of their production and detection is quite similar to
the M$\rm\ddot{o}$ssbauer effect of gamma rays.

Can M$\rm\ddot{o}$ssbauer neutrinos oscillate? An affirmative answer
to this question has been given in Ref. \cite{Lindner}, although
there were some controversies \cite{Bilenky}. Today we have many
technical difficulties in realizing the oscillation of
M$\rm\ddot{o}$ssbauer neutrinos, but in the long run it might not be
impossible to do such a highly-statistical
$\overline{\nu}^{}_e$-disappearance experiment to measure
$\theta^{}_{13}$ with a baseline of some ten meters, to determine
the neutrino mass hierarchy in the absence of terrestrial matter
effects, and to probe new physics beyond the standard picture of
three active neutrino mixing.
\begin{figure*}[t]
\centering
\includegraphics[width=135mm]{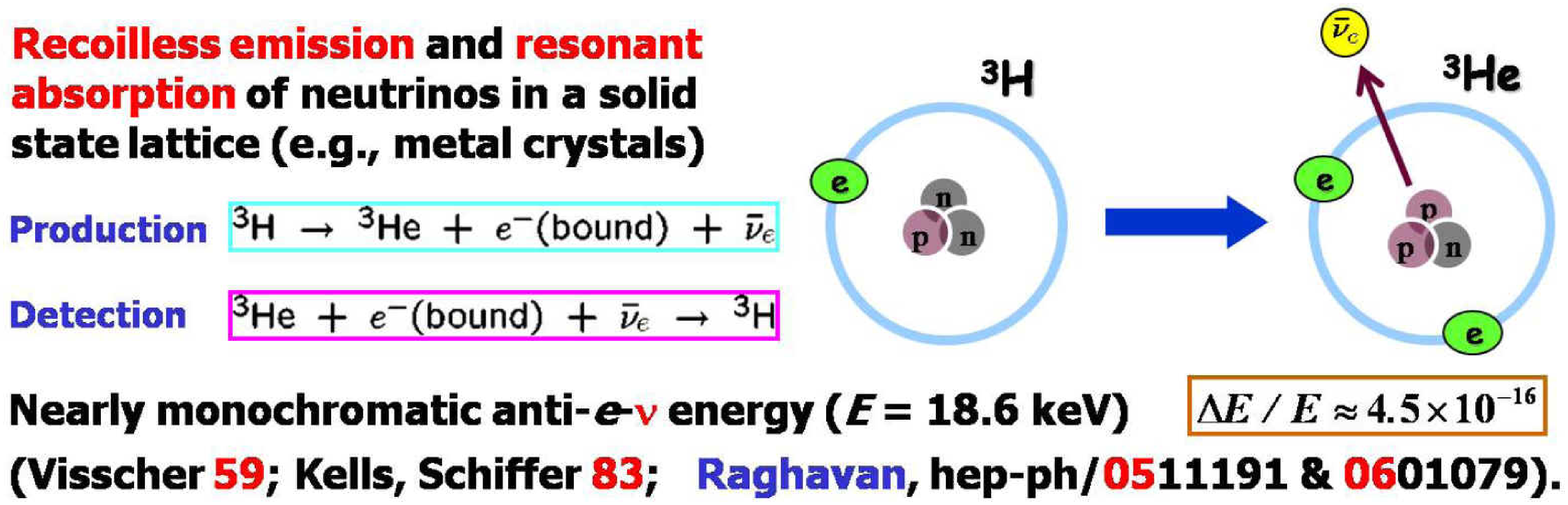}
\caption{An illustration of the recoilless emission and resonant
absorption of $\overline{\nu}^{}_e$ neutrinos.}
\end{figure*}

\subsection{Massive Neutrinos in GUTs and Unparticle Physics}

The study of massive neutrinos in grand unified theories (GUTs)
has been an important topic in theoretical particle physics. In
particular, the type-(I+II) seesaw mechanism can naturally be
realized in the SO(10) GUT with an intriguing unification of
leptons and quarks. Babu's talk in this conference is an excellent
overview of neutrino masses and flavor mixing in a class of GUTs
\cite{Babu}, and thus it allows me to have a good excuse to skip
this topic here.

A brand-new but more or less exotic topic is associated with the
applications of Georgi's unparticle idea \cite{Georgi} to neutrino
physics. A number of authors have considered the influence of
unparticle physics on neutrino decays, neutrino scattering with
electrons, and neutrino oscillations \cite{UP}. Current works on
this subject are purely speculative.

It is a pity that I am unable to cover some other interesting
topics in neutrino physics and neutrino astrophysics. For example,
a lot of works done in the past two years (from ICHEP06 in Moscow
to ICHEP08 in Philadelphia) focus on model building with the help
of supersymmetries, flavor symmetries, GUTs or strings, extra
dimensions, and some new ideas. A possible roadmap of neutrino
mass models can be found in King's talk at {\it Neutrino 2008}
\cite{King}.

\section{THE ROAD AHEAD}

We have known a lot about the properties of three known neutrinos,
but there remain many things that we do not know. The present
situation of model building seems quite messy, although we are more
or less guided by some principles including the self-consistency,
naturalness, simplicity and testability. As argued by Witten
\cite{Witten} at {\it Neutrino 2000}, ``for neutrino masses, the
considerations have always been qualitative, and, despite some
interesting attempts, there has never been a convincing quantitative
model of the neutrino masses". The road ahead is therefore to
establish the {\it unique} and {\it quantitative} theory of neutrino
masses, flavor mixing and CP violation.

In his autobiographic book {\it The Road Ahead} \cite{Gates}, Bill
Gates admits that ``people often overestimate what will happen in
the next two years and underestimate what will happen in ten". My
mild estimate is that new breakthroughs might be possible in the
near future at three frontiers: the energy frontier set by the LHC,
the intensity frontier associated with neutrino experiments; and the
cosmic frontier for the study of dark matter and dark energy. The
overlap of three frontiers is obvious, and it has much to do with
the topics covered in this talk. I hope that new experimental
discoveries may help to lead us to a fundamental neutrino theory. In
particular, I hope that the LHC might tell us something behind three
known neutrinos --- either new particles or new symmetries, or both
of them.

\begin{acknowledgments}
I would like to thank the conference organizers N. Lockyer, J.
Kroll and A.J. Stewart for inviting me to give this talk and for
helping me to get the visa. I am also grateful to my students S.
Zhou, S. Luo, W. Chao and H. Zhang for their great enthusiasm and
patience in assisting me with many surveys of the neutrino
literature and preparations for the PPT file. This work was
supported in part by the National Natural Science Foundation of
China.
\end{acknowledgments}

\end{document}